# A Quality-Aware Voltage Overscaling Framework to Improve the Energy Efficiency and Lifetime of TPUs based on Statistical Error Modeling

**Alireza Senobari[1], Jafar Vafaei[2], Omid Akbari[1], Christian Hochberger[3], Senior Member, IEEE, Muhammad Shafique[4], Senior Member, IEEE**

[1] Department of Electrical and Computer Engineering, Tarbiat Modares University, Tehran 14115-111, Iran
[2] School of Electrical and Computer Engineering, University of Tehran, Tehran 14395-515, Iran
[3] Department of Electrical Engineering, TU Darmstadt, 64289 Darmstadt, Germany
[4] Division of Engineering, New York University Abu Dhabi (NYU AD), Abu Dhabi 129188, United Arab Emirates

Corresponding author: Omid Akbari (e-mail: o.akbari@modares.ac.ir).

**ABSTRACT** Deep neural networks (DNNs) are a type of artificial intelligence models that are inspired by the structure and function of the human brain, designed to process and learn from large amounts of data, making them particularly well-suited for tasks such as image and speech recognition. However, applications of DNNs are experiencing emerging growth due to the deployment of specialized accelerators such as the Google's Tensor Processing Units (TPUs). In large-scale deployments, the energy efficiency of such accelerators may become a critical concern. In the voltage overscaling (VOS) technique, the operating voltage of the system is scaled down beyond the nominal operating voltage, which increases the energy efficiency and lifetime of digital circuits. The VOS technique is usually performed without changing the frequency resulting in timing errors. However, some applications such as multimedia processing, including DNNs, have intrinsic resilience against errors and noise. In this paper, we exploit the inherent resilience of DNNs to propose a quality-aware voltage overscaling framework for TPUs, named X-TPU, which offers higher energy efficiency and lifetime compared to conventional TPUs. The X-TPU framework is composed of two main parts, a modified TPU architecture that supports a runtime voltage overscaling, and a statistical error modeling-based algorithm to determine the voltage of neurons such that the output quality is retained above a given user-defined quality threshold. We synthesized a single-neuron architecture using a 15-nm FinFET technology under various operating voltage levels. Then, we extracted different statistical error models for a neuron corresponding to those voltage levels. Using these models and the proposed algorithm, we determined the appropriate voltage of each neuron (the voltage level of each column of the X-TPU). Results show that running a DNN on X-TPU can achieve 32% energy saving for only 0.6% accuracy loss.

## I. INTRODUCTION

Neural networks are complex, flexible, and adjustable computation models that can be trained to solve complex problems such as speech recognition, computer vision, encryption, and security [1]. However, these computational models require many computing resources having a high demand for power/energy. Specifically, these requirements limit the applications on low-end devices such as edge computers and embedded systems, where there are rigid limitations on the energy budget and computational power [2]. To mitigate these issues, many recent works have proposed different energy-efficient hardware accelerators such as Eyeriss [3] and CREW [4].

The main computation of a neural network model is the multiply-and-accumulate (MAC) operation done millions of times in the inference phase. For more efficient performance of these computations, hardware accelerators designed for deep neural network (DNN) applications integrate hundreds or thousands of MAC units. The tensor processing unit (TPU) developed by Google is a commercially successful example of such accelerators, which is composed of a systolic array-based architecture that significantly reduces power consumption by lowering memory access count. Google has deployed these accelerators in its data centers for large-scale service, where energy efficiency is one of the main concerns, as well as the thermal design power (TDP) constraints [1]. Also, improving the energy efficiency of these devices is highly beneficial, and may allow using them on edge devices and embedded systems for a longer duration.

Approximate computing is a design paradigm that lowers the energy consumption and/or area of inherently error-resilient applications, such as multimedia processing, machine learning, and data mining [5], in which a reasonable number of errors is acceptable. Since DNNs have inherent error resiliency, where the effectiveness of these networks is often described by their loss or accuracy, they are amenable to employing the approximate computing for them.

Researchers have developed many different methods to leverage approximate computing in designs, ranging from the hardware (like logic simplification [2] and voltage overscaling (VOS) [6]) to the software (like loop perforation [7]) layers. In the VOS technique, the operating voltage of the system is scaled down without changing the frequency resulting in timing errors, where path delays are higher than the clock period. Since scaling the operating voltage increases path delays, timing errors occur more often when the distance between nominal and operating voltage increases [8]. Key benefits of using VOS technique for approximation compared to other methods such as LSB truncation, is that this method requires no hardware redesign and can be configured at runtime to perform in nominal voltage (i.e., exact operations). This technique also improves the lifetime of circuits by reducing voltage effects on aging mechanisms [9]. However, one of the disadvantages of the VOS method is that it requires additional circuits such as extra switches and level shifters (in some situations) to manage the operating condition of the circuit [9].

As TPU has different dataflow compared to other accelerators and heavily relies on data reuse (rather than sending data to memory and fetch it back), applying VOS technique on it results in different error behaviors. To show the effect of employing the VOS on the accuracy and power consumption of a processing element (PE) of TPU, we simulated a single PE under different operating voltage levels. The simulation toolchain of this investigation is discussed in Subsection V.A. FIGURE 1 (a) shows the internal structure of a PE of the TPU, composed of a multiplier, an adder, and several registers. FIGURE 1 (b) presents the power consumption decomposition of this PE. Finally, FIGURE 1 (c) shows the error variance and power consumption of a PE for the investigated voltage levels. Note that for the different investigated voltage levels, only the operating voltage of the multiplier of the PE is overscaled. This is due to the large share of the multiplier on the power consumption of the PE (see FIGURE 1 (b)), while it offers a predictable behavior of using the VOS on the overall TPU output quality.

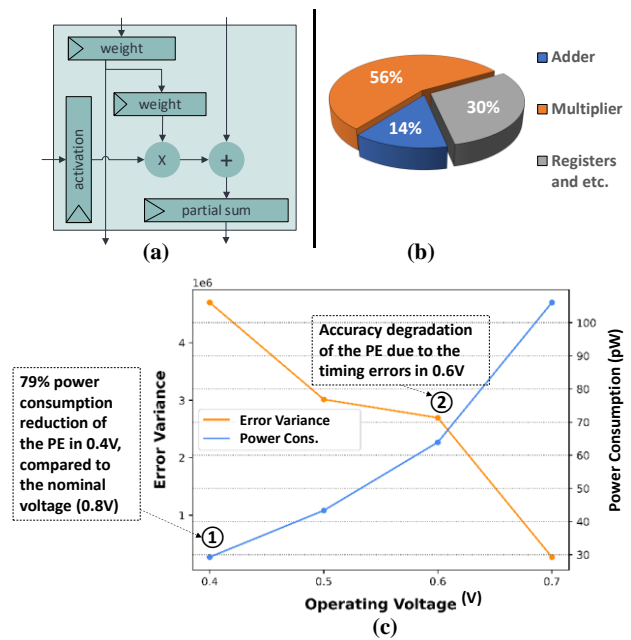

**FIGURE 1. (a)** Internal structure of a processing element (PE) of TPU, **(b)** power consumption decomposition (%) of the components composing a PE, **(c)** accuracy and power consumption of a PE for the different operating voltages.

Based on the results, by overscaling the operating voltage to 0.4 V, where the nominal voltage is 0.8V in the used 15nm FinFET technology, the power consumption of the PE is reduced about 79% (see pointer ① in FIGURE 1). Thus, employing the VOS may provide significant power/energy consumption reduction of TPU. However, employing the VOS may increase the delay of the circuit, and consequently, results in some errors due to the timing constraints violation (e.g., see pointer ② FIGURE 1 (c)). Therefore, to apply the VOS on TPU under a given threshold of accuracy drop, a systematic method is required to determine the voltage levels of the PEs, while the overall accuracy drop is retained upper than the user-defined quality constraint.

In this work, we propose a VOS-based TPU (X-TPU) with the ability of dynamic accuracy configuration to provide various energy and quality of service (QoS) levels. To this end, first, we present the architecture of X-TPU whose PEs supports different voltage levels, as well as a modified weight memory to store the voltage level of PEs corresponding to the required accuracy level. Next, we assign different voltage levels to neural network neurons based on their saliency. To limit the accuracy drop of the accelerator due to induced timing errors, we propose a systematic design method that extracts the error distribution of MAC units under different operating voltages. Then, this information is used to determine the operating voltage of each neuron using an Integer Linear Programming (ILP) method, in a way that the overall quality of the inference phase remains in a user-defined range.

**To summarize**, the main contributions of our work are as follows:

1) Proposing a dynamic quality configurable TPU that leverages VOS to provide various network accuracy levels, as well as different energy consumptions.

2) Developing a statistical quality-aware overscaling framework that leverages an optimization algorithm to determine each neuron's operating voltage such the overall energy consumption of the TPU is minimized, while the output quality is retained above the user-defined one.

3) Investigating the aging effects on the lifetime of PEs of the X-TPU under various voltage levels.

**Paper Organization:** Section II discusses prior works on TPUs and approximate Computing. A brief introduction to DNNs and TPU is noted in Section III, followed by the proposed architecture and framework in Section IV. Section V presents the simulation setup and final results. Finally, this paper is concluded in Section VI.

## II. RELATED WORKS

This section reviews state-of-the-art works on approximate computing, and specifically, those works that employed this computing paradigm for TPUs and machine learning applications.

### A. Approximate Computing

In recent years, approximate computing has received great interest for designing energy- efficient digital systems [10], e.g., leading companies such as Google, Samsung, and IBM employed this computing paradigm in their experimental research and commercial products [11]. Different studies targeting the approximate computing paradigm from the hardware to software layers have been reviewed in [10].

Hardware-level approximation techniques have been conducted in [12][13], primarily targeting DNN applications to increase energy efficiency and performance. In [14], runtime configurable approximate hardware challenges such as error bounding and the required hardware to support the various quality levels have been discussed, and then, a systematic design methodology has been proposed to overcome those challenges for coarse-grained reconfigurable hardware (CGRAs). An approximate CGRA based on the VOS technique has been proposed in [9], such that each PE can operate at different voltage levels. If an application could not leverage errors, all PEs in CGRA operate in exact operation mode (i.e., the nominal voltage level). Otherwise, depending on the tolerable output quality degradation, some PEs can operate on lower voltages to maximize energy efficiency. To determine which PEs can operate on what voltage to optimize energy-accuracy relation, a mapping algorithm based on Integer Linear Programming (ILP) has been proposed that maps DFG nodes to certain PE clusters with known operating voltage. Another approximate reconfigurable architecture for Dadda multipliers has been proposed in [15] based on the VOS technique and LSB truncation.

SRAMs show volatile functionality when using aggressive voltage scaling. To mitigate this issue, [16] proposed a hardware/algorithm co-design methodology that applies VOS to weight SRAMs using memory adaptive training in DNN accelerators. In [6], a new architecture called TE-Drop was proposed for systolic arrays using razor flip-flops. This architecture is designed to detect timing errors that can be caused by low operating voltage in a PE. When an error is detected, the subsequent PE is skipped, and an extra clock cycle is used to correct the erroneous value of the affected PE. The authors also proposed an algorithm that adjusts the voltage under the scaling ratio for each layer of the deep neural network (DNN) to optimize energy consumption and accuracy.

The TE-Drop architecture, initially proposed in [6] as a way to detect and correct timing errors in systolic arrays, was modified in [17] to prevent computation skipping. This modification involved the addition of a transition detector (TD) to each PE. The TD is designed to detect any unexpected transitions that occur near the end of each clock cycle, and sends a warning signal to the next PE if such a transition is detected. Instead of discarding the entire computation of the subsequent PE, only the corrupted partial product is replaced with the correct result from the previous PE. This modification not only helps to improve the accuracy of the system, but also allows for more efficient use of resources by avoiding the need to discard and redo large amounts of computation.

In [12], a method for minimizing the loss of accuracy in a deep neural network (DNN) model when using a particular approximate arithmetic unit was proposed. This method involves the use of a weight replacement function that can be determined for a pre-trained DNN model and a set of available approximate arithmetic units. Given an approximate arithmetic unit as input, the function maps accurate weights to new approximate values, i.e., the function can take a set of accurate weights and, using a particular approximate arithmetic unit, map them to a new set of approximate weights. In [18], a heuristic algorithm was introduced for optimizing the performance of systolic arrays in deep neural networks. This algorithm groups input data sequences of activations based on their induced delay to the output value, and then configures the PEs in a row of the systolic array to operate at an optimistic voltage level for each group of inputs. By doing so, the system is able to operate more efficiently and accurately. In [19], a low-variance reconfigurable multiplier for systolic array matrix multiplication unit (LVRM) was proposed that is capable of operating in three different modes: one exact mode and two approximate modes.

TABLE 1 Comparison of the State-of-the-Art Accuracy Configurable Approximate architectures.

| Ref. | Description | Target Platform | Investigated Application Domain | Objective | HW Modification | VOS | Runtime Configurable |
|---|---|---|---|---|---|---|---|
| [14] | Using logic modification to design an approximate architecture | CGRA | DSP, Multimedia Processing | Energy Efficiency | ✓ | ✗ | ✓ |
| [9] | Using VOS to design an approximate architecture | CGRA | DSP, Multimedia Processing | Energy Efficiency, Lifetime Improvement | ✓ | ✓ | ✓ |
| [15] | Using VOS to design an approximate arithmetic unit | Multiplier | Image Processing, Multimedia Processing | Energy Efficiency, Lifetime Improvement | ✓ | ✓ | ✓ |
| [13] | A neuron-wise voltage assignment and clustering | FPGA | Neural Networks | Energy Efficiency | ✗ | ✓ | ✗ |
| [18] | A heuristic method to analyze input for fault mitigation | Systolic Arrays | Neural Networks | Energy Efficiency | ✓ | ✓ | ✓ |
| [21] | Using VOS on MSB bits of multipliers to achieve an accuracy Configurable NN accelerators | NVDLA | Neural Networks | Energy Efficiency | ✓ | ✓ | ✓ |
| [16] | Applying VOS to weight SRAMs of DNN accelerators | SNNAC | Neural Networks | Energy Efficiency | ✓ | ✓ | ✓ |
| [6] | Per-Layer application of VOS with retraining the NN | TPU | Neural Networks | Energy Efficiency | ✓ | ✓ | ✗ |
| [12] | Selecting the optimized approximate multiplier to compose the layers of NNs | DaDianNao | Neural Networks | Energy Efficiency | ✓ | ✗ | ✗ |
| [8] | Using VOS to design approximate arithmetic units | Multiplier | Image Processing | Energy Efficiency, Lifetime Improvement | ✓ | ✓ | ✓ |
| The proposed X-TPU | A VOS-based neuron-wise voltage assignment using the error models of approximate hardware | TPU | Neural Networks | Energy Efficiency, Lifetime Improvement | ✓ | ✓ | ✓ |

The authors also presented a methodology for determining which operation mode each LVRM should run in based on a heuristic model of errors, taking into account user-defined constraints. This methodology was later applied in [2] and [20].

In [21], VOS has been applied to computing units of the NVIDIA deep learning accelerator (NVDLA), to increase its energy efficiency while retaining a user-defined quality constraint. In [8], a generic accuracy configurable multiplier was proposed, which employs VOS at a block level to reduce the overheads such as control logic and level shifters. A neuron-level voltage scaling approach for DNNs has been proposed in [13] based on the error propagation model at both inter- and intra- layer perspectives. This article assumes that the error rate of the hardware running a DNN has a linear relation with the operating voltage. However, we showed it is not true when applying the VOS technique since the error rate is heuristic when the hardware operates below the nominal voltage. Also, per-neuron voltage assignment has been done through a genetic algorithm. Thus, a clustering algorithm is used to create optimized voltage islands for neurons to lower the area and energy.

### B. Improving the efficiency of TPU
Google first introduced Tensor Processing Unit (TPU) architecture. The main idea of this architecture is to compute the inference phase of a DNN fast and efficiently [1]. Since then, this architecture has been modified in some ways. As an example, [4] proposed a modified TPU architecture with a shared partial product buffer among MACs, which lowers the number of multiplications by using previously computed products.

A new algorithm for compressing sparse matrices was proposed by [22], which utilizes a modified TPU architecture for efficient performance. Although this algorithm offers benefits such as energy efficiency and performance, it may also increase fault rates as technology scales. Furthermore, the widespread use of DNNs and TPUs may pose potential security risks and malware issues. Researchers in [22] showed that only four units with permanent fault among 64k MAC units could drop inference accuracy sharply. They have proposed new techniques called Fault Aware Pruning (FAP) and FAP+, which rely on pruning. The pruned weights are mapped to faulty MACs and then modified TPU to bypass a faulty MAC using a multiplexer. FAP+ also performs a retraining step to improve the pruned DNN's accuracy.

To further overcome threats of hardware trojans and fault injection attacks, [23] presented a toolchain based on the Interactive Proofs (IP) protocol. The prover in this work itself can be described as a two-step computation. The first step is not computationally expensive; thus, it is implemented sequentially in the modified TPU. The second phase is matrix multiplication, which can be performed using the available resources of the baseline TPU (the Matrix Multiplier Unit) to reduce the area overhead of using this protocol.

Table 1 presents a summary of the key features of the state-of-the-art VOS-based architectures. Unlike state-of-the-art works that are not designed for neural networks (such as [9]) or those optimizing power consumption of the DNNs in a layer-wise manner (like [6]), in this paper, we aim at deploying the VOS technique in TPUs where based on the NN architecture used, user can determine each neuron's operating voltage in order to reduce energy consumption while maintaining accuracy at an acceptable range without significant hardware change, retraining or pruning. This also comes with certain additional challenges, e.g., time complexity of finding the best solution and extracting error characteristics of the TPU while running on VOS mode, where we addressed these challenges for the proposed X-TPU in Section IV. In next section, we study the preliminaries required to introduce our proposed X-TPU.

## III. PRELIMINARIES
In this section, first, the aging phenomenon and its mechanisms, as well as the impact of VOS on reducing the aging effects and the delay of circuits are detailed. Then, the DNN model is discussed. Finally, the structure and functionality of TPU are presented.

### A. Aging Mechanisms
By shrinking the technology size in the nanoscale era, aging effects become more important which may affect the system reliability. Aging occurs due to different factors, such as Bias Temperature Instability (BTI), Negative Bias Temperature Instability (NBTI), Hot Carrier Injection (HCI), Time-Dependent Dielectric Breakdown (TDDB), and Electromigration (EM) [24]. These mechanisms may change the characteristics of the transistors, and consequently, impact the performance and reliability of the circuits. Fortunately, reducing the supply voltage can mitigate these destructive effects. Among the aging mechanisms, BTI can directly influence threshold voltage, where the variation of the threshold voltage is obtained by [9]

$$\Delta V_{th,BTI} \cong A e^{\frac{\kappa}{\theta}} t^{\alpha} E_{OX}^{\gamma} f^{\beta} \tag{1}$$

where $A, \kappa, \alpha, \beta,$ and $\gamma$ are technology-dependent constants. Also, $\theta, t, and f$ are temperature, stress time, and duty factor, respectively. Moreover, $E_{OX}$ is the electric field induced on the gate oxide defined by [9]

$$E_{OX} = \frac{V_{DD} - V_{th}}{T_{INV}} \tag{2}$$

where, $T_{INV}$ is the thickness of the inversion layer. Based on (1) and (2), aging-caused change in the threshold voltage (transistor characteristics) is a strong function of the supply voltage, i.e., lowering the operating voltage ($V_{DD}$) reduces the aging effects. In Subsection V.C, (1) is used to evaluate the aging effect on the threshold voltage changes of the VOS-based TPU.

### B. VOS technique
Voltage overscaling reduces the operating voltage of the circuits further than the nominal operating voltage determined by a specific technology scale to achieve higher energy efficiency, as well as an improved lifetime as discussed in the previous subsection. However, narrowing the distance between the operating voltage and threshold voltage increases the delay in the paths of the circuit, where the circuit delay is modeled through the alpha power law defined by [9]

$$Task_{Delay} \propto \frac{V_{DD}}{(V_{DD} - V_{th})^{\alpha}} \tag{3}$$

where $\alpha$ is a technology-dependent constant considered 1.3 for sub-20-nm technologies [9]. Based on (3), using the VOS technique in circuits designed to operate on nominal voltage may violate the timing constraints and induce some timing-related errors in the circuit output. In Section IV, a systematic method is proposed to determine the voltage levels of the PEs of the X-TPU, such that the output accuracy is retained in the range of user-defined quality constraints.

### C. Deep Neural Networks
Deep neural networks (DNNs) are computation models that resemble trees, with a minimum depth of three layers comprising the input layer, hidden layers, and output layer. FIGURE 2 shows structures of two well-known DNNs, including Fully Connected (FC) and Convolutional Neural Network (CNN). In CNNs, each Kernel (filter) corresponds to a neuron in fully connected network. Each node within these layers is made up of multiply-and-accumulate (MAC) operations.

Note that the performance of a neural network on a particular learning task can be influenced by various factors, such as the architecture of the network, the size of the training data, the complexity of the task, the choice of activation functions, and the optimization algorithm used for training. Besides these, the type of learning task (classification or regression) may have a significant influence on the performance of the neural network. In classification tasks, the neural network is trained to predict discrete labels or categories for input data. This type of task typically requires the network to learn complex decision boundaries between different classes. However, neural networks are well-suited for classification tasks because they can learn non-linear relationships between input features and target labels.

In regression tasks, the neural network is trained to predict continuous values for input data. This type of task typically requires the network to learn to map input features to output values. Neural networks can also perform well in regression tasks because they are capable of learning complex patterns and relationships in the data. Overall, neural networks can perform well on both classification and regression tasks, but the specific performance of a network will depend on the specific characteristics of the task and how well the network is designed and trained to handle those characteristics.

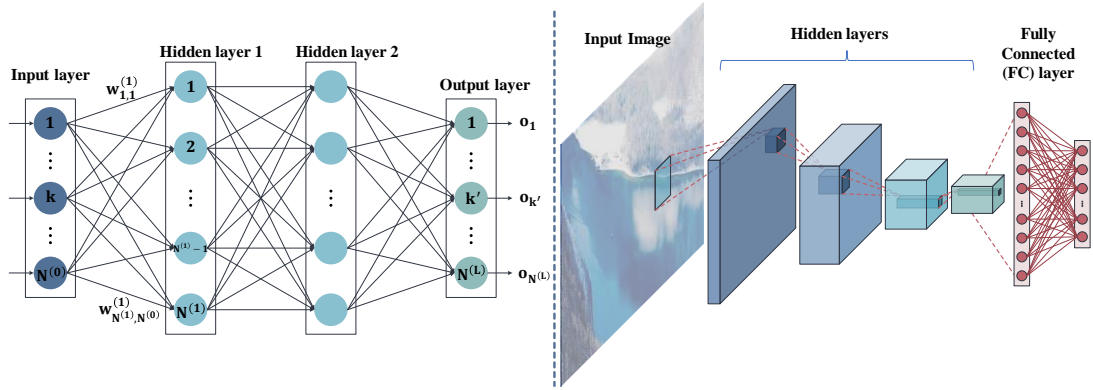

**FIGURE 2.** Two studied DNN architectures, a) fully connected (FC) network and b) convolutional neural network (CNN).

Deploying NN models for use in a service entails two phases: the first phase is the training phase, where the model parameters are adjusted to fit the application, and the second phase is the inference phase, where the inputs are propagated through the layers of the model to reach the output layer. The weights of the model (model's parameters) are updated during the training phase through the backpropagation algorithm. During the inference phase, the output of the $j^{th}$ neuron in layer $i$ is calculated by

$$o_j^{(i)} = g\left(\sum_{k=1}^{n^{(i-1)}} W_{k,j}^{(i)} \cdot o_k^{(i-1)} + b_j^{(i)}\right) \quad (4)$$

where $o_j^{(i)}$ is the output of $j^{th}$ neuron in layer $i$ and $w_{k,j}^{(i)}$ corresponds to the weight from neuron $j$ in layer $(i-1)$ to neuron $k$ in layer $i$. Also, $g(x)$ is the activation function of the layer, where some of the most used activation functions are *sigmoid*, *softmax*, *ReLU*, and *linear*.

The quality of DNNs is represented in terms of *accuracy* and *loss*, where the *accuracy* is only applicable when the network is a *classifier*. Also, the most common loss functions used in DNNs are mean absolute error (MAE), mean-squared error (MSE), mean relative error distance (MRED), and cross-entropy (CE), defined by [25]

$$MAE = \frac{\sum_{i=1}^{n^{(L)}} \left| t_i - o_i^{(L)} \right|}{n^{(L)}} \quad (5)$$

$$MSE = \frac{\sum_{i=1}^{n^{(L)}} \left( t_i - o_i^{(L)} \right)^2}{n^{(L)}} \quad (6)$$

$$MRED = \frac{\sum_{i=1}^{n^{(L)}} \left| \frac{t_i - o_i^{(L)}}{t_i - o_t^{(L)}} \right|}{n^{(L)}} \quad (7)$$

$$CE = -\sum_{c=1}^{n^{(L)}} C_c \, log\left(o_c^{(L)}\right) \quad (8)$$

where $t$ and $o$ are the target and output values, respectively. Also, $L$ is the network's last layer, and $C$ is the class number to which that specific output belongs.

### D. Tensor Processing Unit
During the inference phase, DNNs primarily perform matrix multiplications, which are a sequence of MAC operations. However, these matrix multiplications are computationally intensive and require significant resources in terms of clock cycles and memory operations. Systolic arrays are an effective solution to optimize those operations and take advantage of data reusability, as they have been shown to provide significant benefits [1]. The architecture of TPU is shown in FIGURE 3.

TPU, at its core, is a systolic array of MAC units formed as a 2D grid that manages the data flow in a way that data is being fetched from memory (weight or activation) in multiple clock cycles. TPUs can be designed to operate in weight stationery (WS) or output stationary (RS) modes. In WS mode, weights are prefetched from weight memory into intermediate weight registers in each PE. Then activation values stream from left to right. In the systolic array architecture, partial sums are calculated by each PE and then passed to the next PE in a cascading manner. This allows for the efficient use of parallel processing as the calculation of one PE contributes to the next.

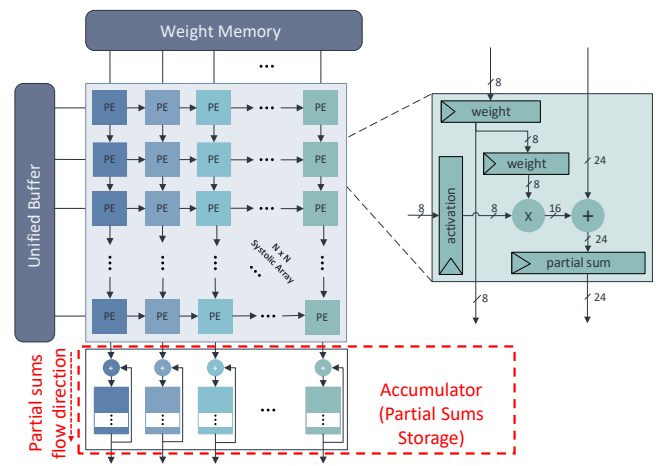

**FIGURE 3.** Architecture of TPU

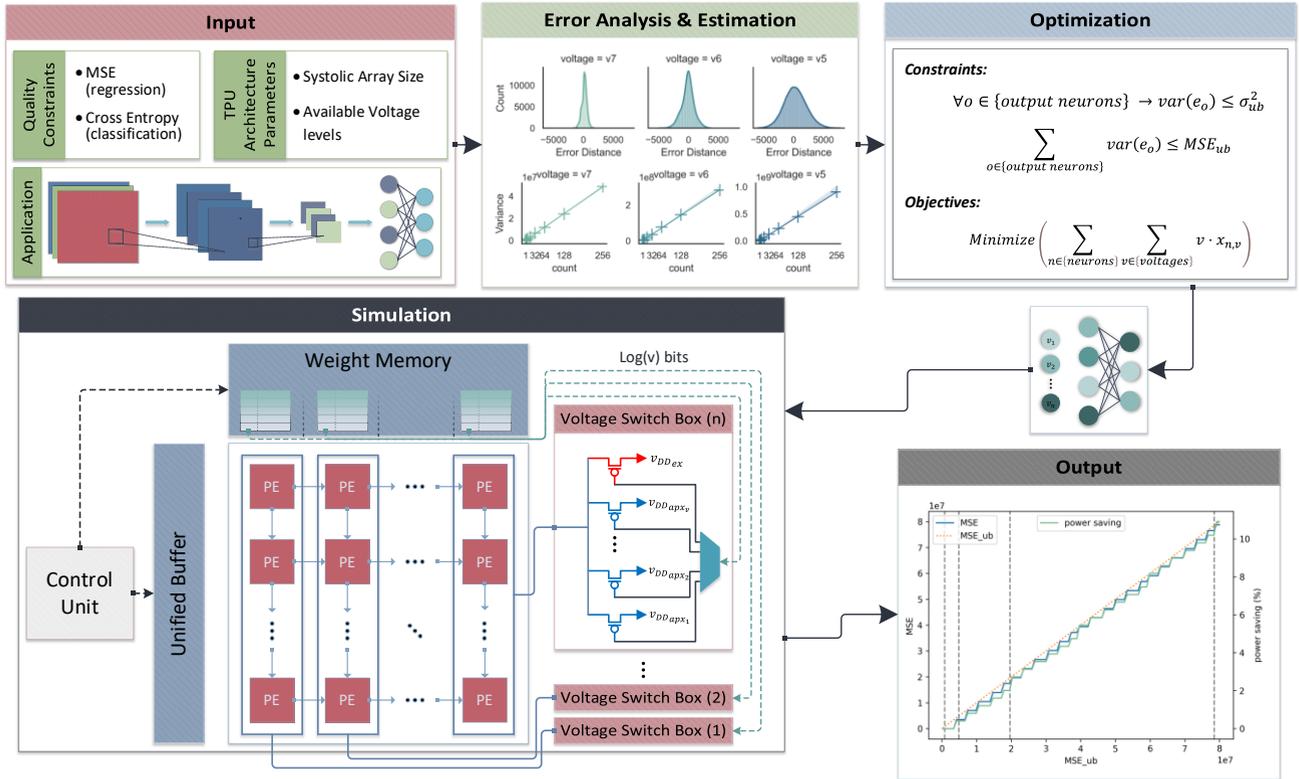

**FIGURE 4.** System overview of the proposed X-TPU framework

Each PE located in the same column of the array shares the same input data, and the results obtained from a PE are passed down to the next PE located in the same column. Thus, when a PE completes its calculations, it passes on the partial sums obtained to the PE located below it in the same column, enabling a smooth and seamless flow of data and computation throughout the array. This cascading effect of computation helps to speed up the overall processing time and leads to more efficient use of the system's resources.

For a TPU with a systolic array with the size of $N \times N$, the first results of the multiplication of a vector with $n$ elements are ready after $n$ clock cycles ($2n$ if the weight prefetch phase is included), and the complete multiplication result is calculated after $2n$ cycles ($3n$ if the weight prefetch phase is included). Both weights and activations stream into the systolic array at once with one clock delay in RS mode. The result of each MAC unit is accumulated in the output register of that unit. If the multiplicative arrays are larger than the systolic array, the partial sums are collected in an accumulator unit, waiting for the remaining results.

## IV. The Proposed VOS-based Approximate TPU (X-TPU)

First, we present a quick overview of our proposed runtime quality adjustable VOS-based TPU (X-TPU) architecture. Then, we describe our proposed framework to utilize the capabilities of X-TPU to minimize energy consumption while retaining the user-defined accuracy bounds. FIGURE 4 provides an overview of our proposed methodology, illustrating the flow from the user inputs to optimized parameters and the validation output. In this designed system, the inputs are: user-defined quality constraints, available architecture parameters (i.e., systolic array size and available voltage levels), and the application, for instance, a neural network model that should run on TPU. These inputs are then passed to the next stages for error analysis and evaluations. Hardware parameters and characteristics are passed to an architecture simulator to analyze the impact of voltage overscaling on the output. After completing this step, which involves extracting the saliency of neurons from the NN model parameters, the results are then passed to an optimization procedure.

In this procedure, by selecting the appropriate voltage level of each neuron, the energy consumption of the TPU is minimized, while the NN output quality is retained above the user-defined quality threshold. The output of this optimization step consists of a set of <neuron, voltage> tuples. These tuples are encoded and added to the model's weights, and then, during runtime on the hardware, these encoded values determine the operating voltage of each neuron. Finally, the output quality of the NN is assessed. In the following, the proposed X-TPU architecture and the method of selecting the operating voltage level of neurons are detailed.

### A. The Proposed X-TPU Architecture

The baseline TPU performs the forward pass (inference operation) on an 8-bit fixed-point quantized pre-trained DNN model that has parameters (weights) varying from -128 to 127. Lower values have less impact on the accuracy of the model output. For example, zero values can be trimmed off from the model without losing any information in the calculations [26]. As an example, FIGURE 5 shows the distribution of weight values in a simple FC network with a hidden layer composed of 128 neurons, and an output layer consisting of 10 neurons, trained on the MNIST dataset.

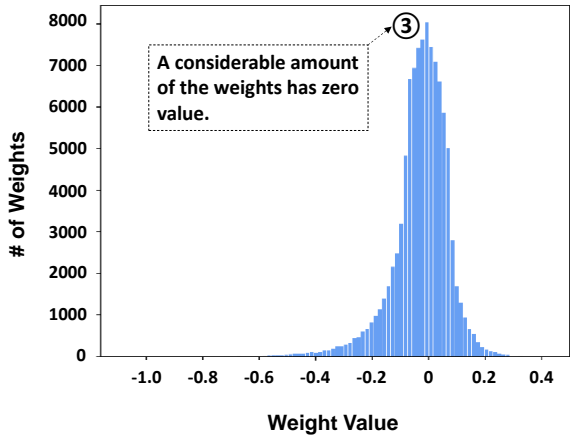

**FIGURE 5.** Distribution of weight values in a simple FC neural network with one hidden layer of 128 neurons and the output layer of 10 neurons trained on the MNIST dataset.

As shown in this figure, a considerable amount of the weights has zero value (see pointer ③ in FIGURE 5). Considering that each weight in the model occupies one PE in the TPU, zero value weights, or the ones with less impact on the output neurons (i.e., the non-important neurons) may waste energy. In the case of sparse arrays, this wasted energy is dominant [27].

X-TPU provides mechanisms to assign lower operating voltages to such non-important neurons to eliminate this excess power usage. However, the baseline TPU is not capable of applying different voltages to different PEs of the MXU. Therefore, it is required to modify the baseline architecture to support VOS. Also, the modified architecture should be flexible to model change (the model which runs on the TPU).

As mentioned in Section I, in the proposed X-TPU, the VOS technique is applied to the multipliers of PEs according to FIGURE 6 (a). Because, as shown in FIGURE 1 (b), a large share of power consumption of the PE (about 56%) belongs to the multiplier, as well as applying the VOS only to the multipliers provides a predictable accuracy drop on the TPU output quality. In detail, applying the overscaled voltage to the whole PE can cause the error of one PE to propagate and accumulate with other PE's errors, as PEs are chained together. This chained accumulation may result in an erratically huge output error, which makes the error modeling of the X-TPU more difficult. Furthermore, this accumulation induces a correlation between errors of PEs. Overall, in order to eliminate these correlations and accumulation of errors, the VOS technique is only applied to the multiplier of PEs.

Note that each column in the TPU represents a neuron in a fully connected network or a kernel in a CNN. Thus, the runtime voltage selection is made possible by appending selection bits to the MSB bits of the weights stored in the weight buffer. Also, the number of selection bits depends on the total number of provided voltage levels '$v_n$'. As an example, to support four voltage levels, including three overscaled (approximate) voltage levels and a nominal (exact) one, two selection bits are added to the weight register. Also, as shown in see FIGURE 6 (a), by applying the VOS to the multipliers, each PE includes two regions with different operating voltages: the approximate region with a low operating voltage ($V_{DD\_Low}$), including the multiplier, and the exact region with a high operating voltage ($V_{DD\_High}$), composed of the other components of the PE such as adder and registers. Note that, for driving the components operating in the $V_{DD\_High}$ (i.e., the adder) by the low voltage components (i.e., the multiplier), level shifters should be employed at the outputs of low voltage components. In detail, level shifters match the operating voltage of outputs of the approximate region to the exact voltage level. Schematic of the used level shifter is shown in FIGURE 6 (b) by the *LS* component, where its internal circuit is represented in FIGURE 6 (c).

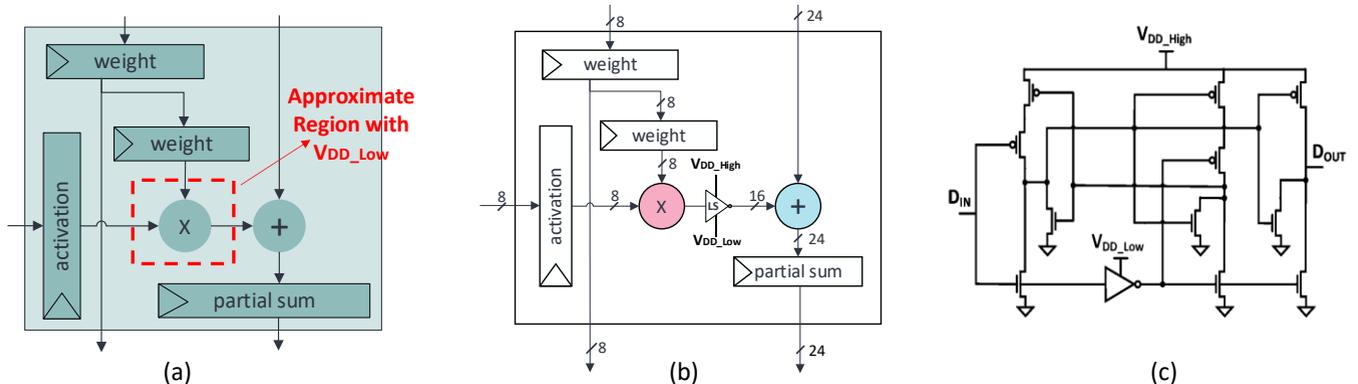

**FIGURE 6.** (a) The approximate region of the PE as being applied in simulations, (b) level shifters added to the PE, (c) level shifter circuit diagram

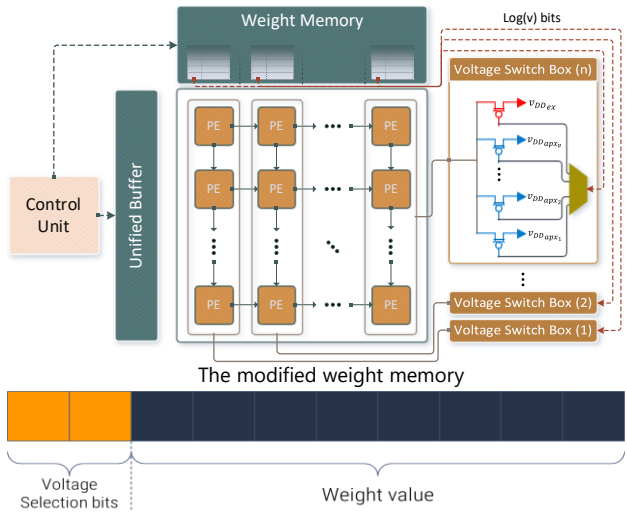

**FIGURE 7. X-TPU architecture and the modified weight memory bit representation to support voltage selection using the voltage selection bits.**

FIGURE 7 shows the architecture of the X-TPU that includes the components required to provide the VOS-based runtime adjustable quality, including voltage switch boxes that are added to each column of the TPU, the bit schema of the weight memory that is designed to operate with voltage switch boxes, and also level shifters. Thus, X-TPU is designed to be capable of applying different voltages to different columns of the PEs as well as the flexibility of voltage mapping (reconfigurability) through voltage selection via voltage selection bits embedded in weight memory.

Finally, to utilize the runtime quality adjustable feature of the proposed X-TPU architecture, it is necessary to establish a method for translating user-defined quality degradation constraints into the voltage levels of neurons, specifically the voltage levels of columns in the X-TPU array. It should be noted that this translation process is carried out while minimizing energy consumption. Thus, in the next subsection, we extract a statistical error model for a PE under various voltage levels, considering the voltage-induced errors.

By utilizing these extracted models and determining the sensitivity of each neuron to the output quality, we reach a systematic design method that enables us to determine the operating voltage of each neuron, ensuring that the network's output quality remains within an acceptable range.

### B. Statistical Error Modeling of Processing Elements

To comprehend the implications of employing the VOS technique on the X-TPU, it is crucial to evaluate the effect of different operating voltages applied to each column of the X-TPU on the magnitude of errors at the DNN's output. Applying the VOS exclusively to the multiplier, eliminates error correlation between PEs. The computations performed in each column of the TPU can be represented by:

$$O_c = \sum_{i=1}^{k_c} W_{c,i} \cdot A_i \qquad (9)$$

where $O_c$ is the output of column $c$. $W_{c,i}$ and $A_i$ are weight and activation values of $i^{th}$ PE in column, respectively, and also, $k$ is the number of PEs in a column. By considering the operation errors, (9) can be written as:

$$O_c + e_c = \sum_{i=1}^{k_c} (W_{c,i} \cdot A_i + e_{c,i}) \qquad (10)$$

where $e_{c,i}$ and $e_c$ are error of $i$th PE in column $c$ and total error generated in column $c$, respectively. As mentioned in Subsection III.B, operating at a voltage lower than the nominal one may introduce timing errors.

According to the simulation results presented in Subsection V.B, these errors exhibit a normal distribution when the VOS is applied only to the multiplier of PE. Given this assumption and considering that the errors in each PE are not correlated with the errors in the subsequent PEs ($cov(e_{PE_i}, e_{PE_{i+1}}) = 0$), and by using (10), we can formulate the error generated in each column by:

$$e_c = \sum_{i=1}^{k} e_i = k \cdot e \qquad (11)$$

where $e$ is a normally distributed random variable corresponding to the output error of each PE. Also, $k$ is a constant. Thus, the expectation and variance of the errors can be calculated as:

$$E(e_c) = k \cdot E(e) \qquad (12)$$

$$Var(e_c) = k \cdot Var(e) \qquad (13)$$

In the next subsection, these equations are used to calculate the error sensitivity (*ES*) of neurons.

### C. Error Sensitivity of Neurons

Each neuron in a network has a varying impact on the output results, depending on its weight values. Furthermore, the architecture of the neural network model also affects the saliency of the neurons [26]. To determine this saliency, first, the *ES* of all neurons is computed by injecting noise based on statistical data derived from equations (12) and (13) into each neuron of the trained model. Subsequently, the ES of the output in relation to each neuron can be calculated by:

$$ES_i^{(l)} = \frac{MED_{net}}{e_i^{(l)}} = \frac{MED_{net}}{k_i^{(l)} \cdot e} \qquad (14)$$

where $ES_i^{(l)}$ is the error sensitivity of $i^{th}$ neuron in layer $l$, and $k_i^{(l)}$ is the number of PEs in neuron $i$ of layer $l$.

It is worthy to note that the forward pass equation shown in (4) can also be used to calculate the *ES* of the neurons to prevent heavy computations of the (14), i.e., by using (4) and

(10), and for a fully connected model, the error of each output neuron can be calculated by:

$$o_i^{(L)} + e_{o_i} = f\left(w_i^{(L)} \cdot g(w^{(L-1)} \cdot \ldots \cdot p(w^{(0)} \cdot x + e^{(0)}) + \cdots + e^{(L-1)}) + e_i^{(L)}\right) \quad (15)$$

$$e_{o_i} = f\left(w_i^{(L)} \cdot g(w^{(L-1)} \cdot \ldots \cdot p(e^{(0)}) + \cdots + e^{(L-1)}) + e_i^{(L)}\right) \quad (16)$$

Then, the *ES* of each neuron can be evaluated by:

$$ES_{ji}^{(l)} = \frac{e_{o_i}}{k_j^{(l)} \cdot e} \quad (17)$$

where $i$ is the index of the output neuron, and $j$ is the index of the intended neuron in layer $l$. The calculated *ES* can be effectively utilized to identify the neurons within the NN model that have lesser impact on the error. Consequently, these neurons can be selected for operating on overscaled voltage. The *ES* can also be used for determining which neurons can be pruned, as well as identifying any redundant or unnecessary neurons for a given task.

### D. Voltage Assignment of Neurons
After determining the *ES* of neurons for the different operating voltages with the proposed formulas in the previous subsection, now, the operating voltage of all neurons should be selected. This involves determining the operating voltage of neurons in a manner that minimizes the total energy consumption while still satisfying the user-defined quality constraint. However, exhaustively exploring all possible combinations of voltages is an NP-Complete problem, making it infeasible, especially for models with a large number of neurons [6]. To address this challenge, we propose an integer linear programming (ILP) based approach, which guarantees to find the optimum solution, rather than the other optimization algorithms like evolutionary algorithms (e.g., genetic algorithm) that cannot guarantee the optimal solution for the zero/one problems [28]. With the ILP approach, optimization is achieved by defining constraints that consider quality requirements, along with an objective function. In our case, the objective is to minimize the energy consumption of the TPU while ensuring that the loss value remains at the user-defined level, as follows:

*Objective:*

$$Minimize \sum_{n \in \{neurons\}} E_n \quad (18)$$

*Constraint:*

$$Q_o < 1 - Q_{D\_UB} \quad (19)$$

where $E_n$ is the energy consumption of the neuron $n$. Also, $Q_o$ and $Q_{D\_UB}$ are the output quality and user-defined output quality degradation upper bound, respectively, which are numbers between 0 (lowest quality) and 1 (highest quality).

Assuming $v$ different voltage levels are supported by the X-TPU, to force each neuron to operate in only one voltage level, the following constraint is declared:

$$\forall n \in \{neurons\}, \sum_{v \in \{available\ voltages\}} x_{n,v} = 1 \quad (20)$$

where $x_{n,v}$ is a binary variable, which is set to 1 when the neuron $n$ is operating in voltage $v$. Note that $x_{n,v}$ is a member of set $X$, where $X$ is defined as

$$X = \left\{x_{n,v} | n \in \{neurons\}, v \in \{available\ voltages\}\right\} \quad (21)$$

The energy consumption of a digital circuit is directly proportional to the square of the operating voltage, denoted as $E \propto v_{DD}^2$ [29]. Consequently, to minimize energy consumption, it is advantageous to minimize the operating voltage of the submodules within the circuit. Accordingly, the objective function can be formulated as the minimization of the sum of the operating voltages of all neurons, considering that each module can operate at different voltage levels. Thus, in our case, the objective function can be expressed as follows:

$$minimize \sum_{n \in \{neurons\}} \sum_{v \in \{voltages\}} x_{n,v} \cdot v \quad (22)$$

The quality of a neural network can be described by providing the accuracy for classification purposes, or the loss value of the network obtained from various loss functions such as MSE. To this end, (6) can be rewritten as:

$$MSE = \frac{1}{n} \sum_{i \in \{output\ neurons\}} e_i^2 = \frac{1}{n} \sum (y_i - \hat{y}_i)^2 \quad (23)$$

where $y_i$ and $\hat{y}_i$ are predicted and true output for the $i^{th}$ neuron, respectively. Also, $e_i$ is the error for the $i^{th}$ output neuron. In the case $E(e_i) = 0$, for example when the occurrence and intensity of error is stochastic with normal-like distribution and zero bias, MSE is equivalent to the variance of the output errors:

$$var(e) = \frac{1}{n-1} \sum (e_i - \overline{e_i})^2 \xrightarrow{\overline{e_i} = 0} \frac{1}{n} \sum e_i^2 \quad (24)$$

Note that in our work, since 1,000,000 uniform random numbers (instead of all possible inputs which were too many) were used, in eq. (24), we have used *(n − 1)* instead of *n*. In these cases, for calculating the variance, instead of *n*, *(n − 1)* is used in the denominator, which this modification is called Bessel's correction [34].

In detail, the equation $E(e_i) = 0$ holds true when the expected value of the error for the $i^{th}$ output neuron is zero. This implies that, on average, the predictions made by the neural network for the $i^{th}$ output neuron are equal to the true values. This condition is typically satisfied when the neural network is well-calibrated and has been trained on a sufficiently large and representative dataset.

The error variance of the NN model is calculated by:

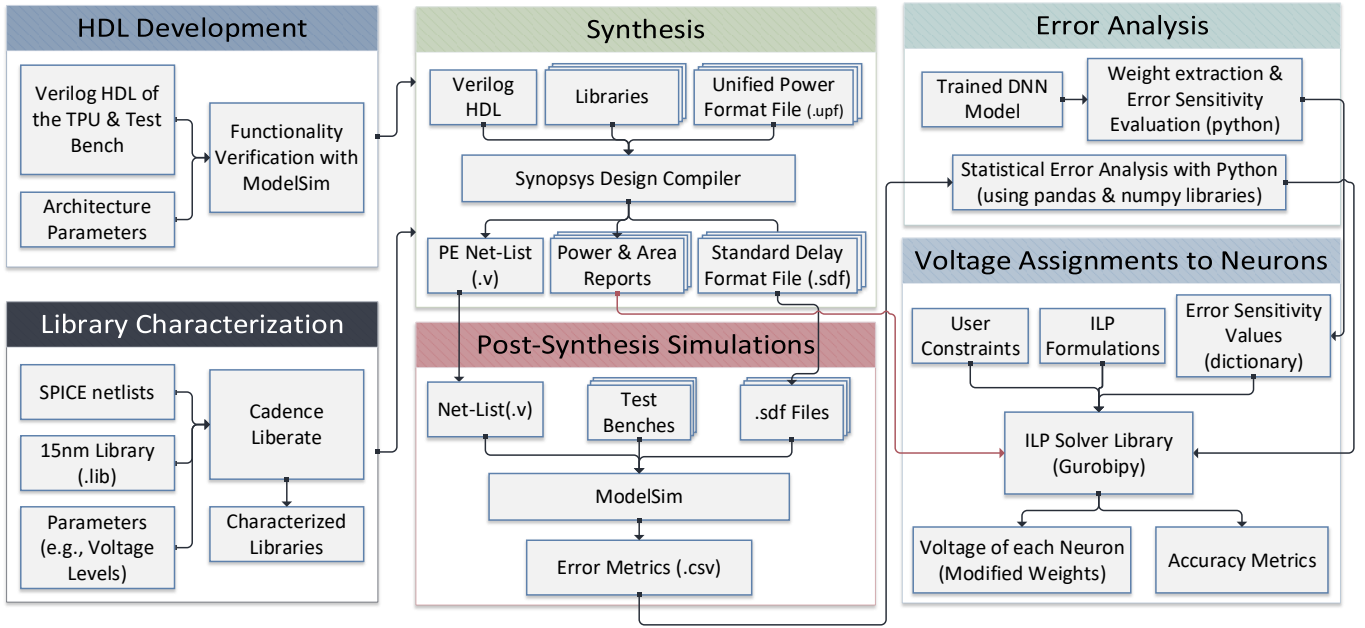

**FIGURE 8.** X-TPU simulation flow. The output of this flow is the set of (neuron, voltage) values for all neurons, as well as estimated (expected) accuracy and loss values.

$$Var(net) = \frac{\sum_{i=1}^{n^{(L)}} \left( \hat{o}_i^{(L)} - \bar{o} \right)^2}{n^{(L)} - 1} \qquad (25)$$

where $\hat{o}_i^{(L)}$ and $\bar{o}$ are erroneous output and desired output, respectively. Also, $L$ is the total number layers in the NN model. Assuming the voltage-induced errors in the model ($\hat{o}_i^{(L)} = o_i^{(L)} + e_{o_i}$), (25) can be rewritten as:

$$Var(e_{net}) = \frac{\sum_{i=1}^{n^{(L)}} e_{o_i} \left( e_{o_i} + 2 \left( o_i^{(L)} - \bar{o} \right) \right)}{n^{(L)} - 1} \qquad (26)$$

Finally, we use the variance of the output neurons as a quality metric of the neural network model. Thus, based on the statistical error model of the X-TPU (see Eq. (17)), the quality constraint shown in (19) can be formulated as follows:

$$\sum_{o \in \{output\ neurons\}} \sum_{v \in \{voltages\}} \sum_{n \in \{neurons\}} var(ES_n \cdot e_n) \cdot x_{n,v} < MSE_{UB} \qquad (27)$$

$$\sum_{o \in \{output\ neurons\}} \sum_{v \in \{voltages\}} \sum_{n \in \{neurons\}} ES_n^2 \cdot var(e_n)_v \cdot x_{n,v} < MSE_{UB} \qquad (28)$$

$$\sum_{o \in \{output\ neurons\}} \sum_{v \in \{voltages\}} \sum_{n \in \{neurons\}} ES_n^2 \cdot k_n \cdot var(e_n)_v \cdot x_{n,v} < MSE_{UB} \qquad (29)$$

where $MSE_{UB}$ is the user-defined MSE increment upper bound ($Q_{D\_UB}$). It is worth noting that in neural network models which use linear activation functions, $ES$ can be replaced by the corresponding L2 norm of the neuron's weights denoted by $\|W\|_2$ in (29). By solving the ILP formulations presented in equations (20), (22), and (29), the voltage level of each neuron within the neural network model is determined. This is achieved by obtaining the variable $x_{n,v}$ for each neuron. Subsequently, the set of voltages corresponding to each neuron is embedded into the model's weights, as described in

Subsection IV.A. These embedded voltage values then determine the operating voltage of each column (PE) within the X-TPU. The selection of the operating voltage for each column is based on the tolerable quality loss ($Q_{D\_UB}$), allowing for on-the-fly adjustments.

## V. RESULT AND DISCUSSION
In this section, first, the simulation setup for evaluating the proposed framework is discussed. Then, error distributions of the PEs for the various voltage levels are analyzed, and also, the X-TPU framework is examined for the MNIST dataset. Finally, the aging effect on the X-TPU architecture is studied.

### A. Simulation Setup
FIGURE 8 shows the developed tool flow and simulation setup for X-TPU synthesis and accuracy analysis. The X-TPU architecture was implemented hierarchically using Verilog HDL. The HDL codes were compiled and verified using ModelSim. Also, to characterize the behavior of the X-TPU architecture under different overscaled operating voltages (i.e., 0.5V, 0.6V, and 0.7V), different technology libraries corresponding to those voltages levels are generated based on the 15-nm FinFET Open Cell Library (OCL) technology [30] and by using Cadence Liberate [31]. Note that the used 15-nm FinFET library with the nominal voltage (i.e., 0.8V) is used for the non-overscaled voltage level (i.e., the exact operating mode). Furthermore, unified power format (UPF) files were generated for each approximate voltage level, considering that each PE has both exact and approximate voltage regions. These UPF files were utilized in the synthesis process with the Synopsis Design Compiler (DC) to produce netlists, power and area reports, as well as standard delay format (SDF) files for each voltage level.

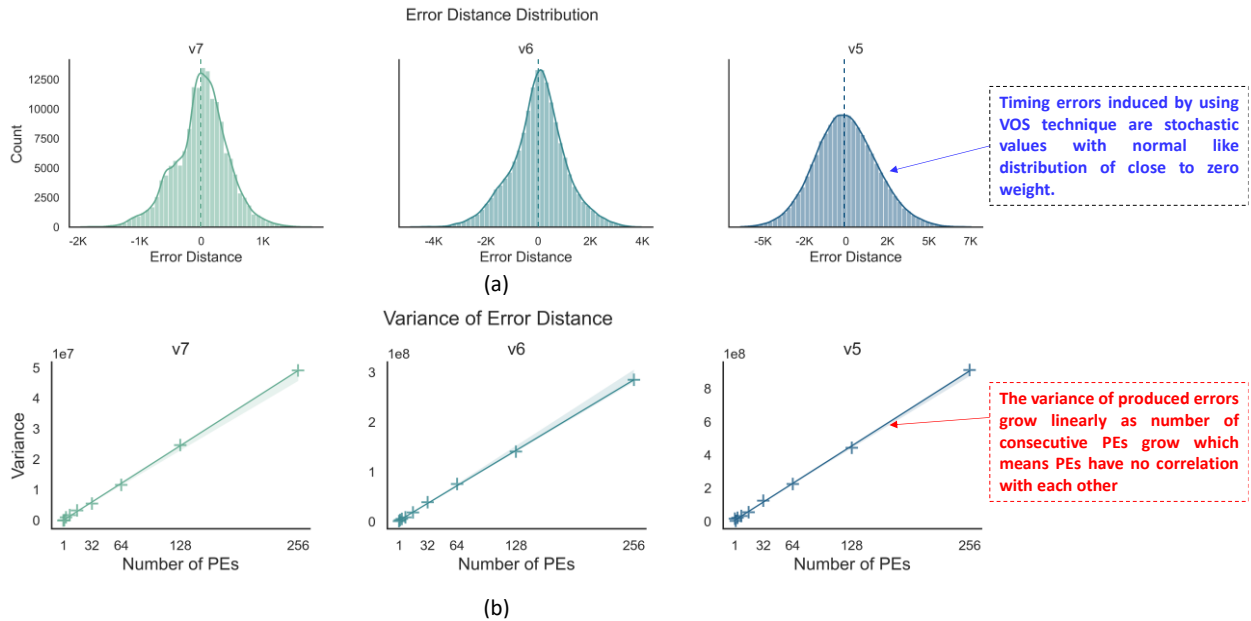

**FIGURE 9.** Statistical analysis results of TPU column. (a) Error Distribution of a single processing element operating on different voltages, (b) effect of TPU column size on the variance of error in different operating voltage

To investigate the error correlation between PEs and obtain a reliable estimate of the error distribution when operating at overscaled voltages, PE columns consisting of 1 to 256 PEs were simulated. For this target, a TCL script was developed to automate the synthesis process under the different parameters' variations. The synthesized netlists, along with the corresponding SDF files, were then subjected to post-synthesis simulation and static timing analysis using the ModelSim simulator. Post-synthesis results were collected in CSV files and analyzed using a custom Python script.

To determine the voltage levels of each neuron, the statistical error model, along with the relevant formulations, and also DNN model parameters, were fed into the Gurobipy ILP solver to solve equations (20), (22), and (29).

Note that the ILP is an NP-complete problem, where the required time to find the solution is proportional to the numbers of constraints and variables (i.e., number of available voltages and neurons in equations (20), (22), and (29)). Based on the simulations, the used ILP solver (i.e., Gurobipy) found the solutions in less than 54.7 seconds. However, one may use heuristic approaches to solve the voltage assignment problem of X-TPU and determine the voltage level of each neuron within the neural network model, in the cases that the solution time of the ILP problem becomes too much.

It should be noted that a 128×10 FC DNN network trained on the MNIST dataset was used in this study. As explained in Subsection IV.A, the obtained voltage levels for the neurons were embedded in the weight values stored in the weight memory. Also, a Python script was developed to generate the augmented weights (see FIGURE 7), by appending the proper voltage selection bits based on the obtained voltage levels, which were then employed in the X-TPU testbench for the final verification. Note that due to the computational demands and memory requirements associated with simulating an entire FC or CNN model in ModelSim, particularly with SDF annotations, it was not feasible to conduct such simulations within a reasonable timeframe.

Specifically, timing analysis simulations for the whole NN model implemented on the TPU hardware is very computing intensive, e.g., as studied in [6], simulation of a LeNet model with a single input on a TPU with the array size of $256 \times 256$ using a 40-core Intel Xeon server machine took 8 hours. To overcome this challenge, the solution of in [6], was to simulate 32 randomly selected column which is a Matrix Multiplication (MM) operation in TPU and extract error probabilities from it and inject timing errors to whole NN model. Similarly, our focus in the verification process was on the MM operation. Specifically, the proposed framework was verified using a 16×16 MM testbench.

**B. Error Distribution of PEs**
Error distribution of a single PE under different operating voltages has been acquired by simulations over one million random inputs fed into columns of PEs with different sizes. The distribution variances obtained from this simulation are listed in Table 2. Also, the variance distribution plots for each voltage level have been shown in FIGURE 9 (a). Moreover, the variance values with respect to the number of PEs are shown in FIGURE 9 (b). These values are used to calculate (24). The simulation results of the considered 16×16 MM benchmark have been plotted in FIGURE 10, in which the primary (secondary) vertical axis shows calculated MSE (power saving) for the different values of $MSE_{D\_UB}$. Specifically, in this figure, the orange dotted line shows the MSE limit determined by the user, i.e., the simulated MSE below this line met the user constraint.

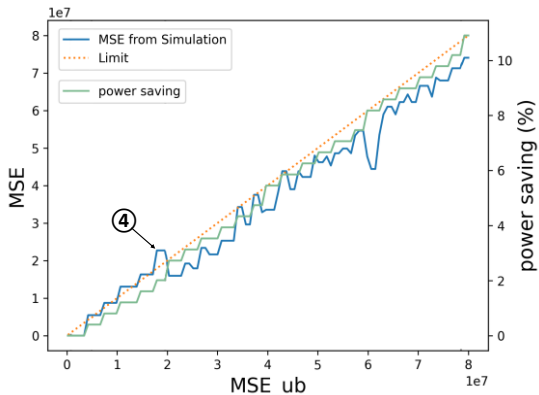

**FIGURE 10.** Simulated MSE values for the different MSE increase upper bounds, along with the power saving of the X-TPU.

TABLE 2 Statistical parameters for the different sets of PEs () inside the X-TPU, correspond to the various number of PEs in each column of the different sizes of X-TPU.

| | VARIANCE | | |
|---|---|---|---|
| **VOLTAGE LEVEL** | **0.5V** | **0.6V** | **0.7V** |
| **NUMBER OF PES** | | | |
| 1 | 3.0E+06 | 1.4E+05 | 2.0E+05 |
| 2 | 1.9E+07 | 3.0E+06 | 7.5E+05 |
| 4 | 1.0E+07 | 3.2E+06 | 8.5E+05 |
| 8 | 2.8E+07 | 8.2E+06 | 9.1E+05 |
| 16 | 6.0E+07 | 1.9E+07 | 2.9E+06 |
| 32 | 1.1E+08 | 3.4E+07 | 5.5E+06 |
| 64 | 2.3E+08 | 7.2E+07 | 1.3E+07 |
| 128 | 4.5E+08 | 1.4E+08 | 2.5E+07 |
| 256 | 8.9E+08 | 2.9E+08 | 4.9E+07 |

As shown in this figure, in most cases, the simulated MSE is close to or below the use-defined constraint. However, in some cases (e.g., see pointer ④ in FIGURE 10) the $MSE_{UB}$ was violated that is due to the intrinsic error of Eq. (23) [25][32]. In total, the quality constraint violations are negligible, which is on average, 0.3% for all the different studied $MSE_{D\_UB}$. Moreover, for the considered $MSE_{UB}$s, power consumption of the X-TPU was decreased between 0 and 12%.

FIGURE 11 shows the calculated $ES$ obtained by (14), for neurons of a 128×10 fully connected DNN trained on the MNIST dataset. As shown in this figure, neurons of the hidden layer have lower $ES$ (all less than 0.4) compared to the output layer's neurons, where the $ES$ of the output layer is almost 1, i.e., the neurons of the hidden layer have less impact on output accuracy and can be proper candidates for applying the VOS. Afterwards, the acquired $ES$ of each neuron, along with the proposed ILP formulation in (20)~(29), are fed to the used ILP solver (GurobiPy) for selecting the operating voltage levels of neurons of the simulated neural network. Specifically, by applying our proposed framework on a fully connected DNN trained on the MNIST dataset using Linear activation function, for the MSE increment upper bounds ranging from 1% to 1000%, the assigned voltage levels of all neurons were shown in FIGURE 12.

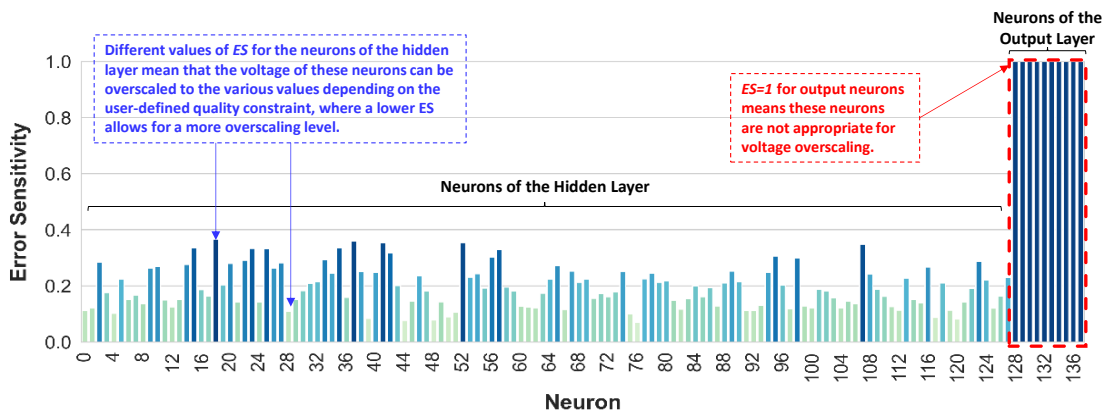

**FIGURE 11.** Error Sensitivities of all neurons of a 128×10 FC network trained on the MNIST dataset.

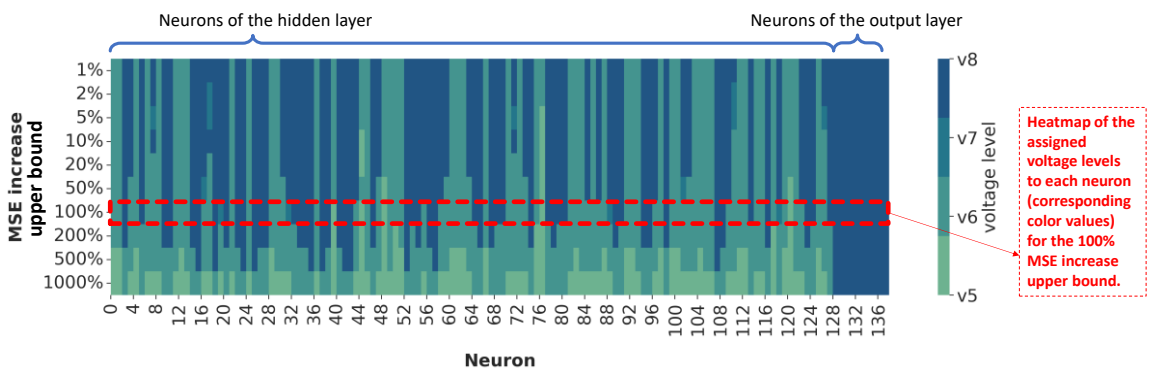

**FIGURE 12.** The assigned voltage level of all neurons for a 128×10 FC DNN under the different MSE increase upper bounds, visualized using a heatmap.

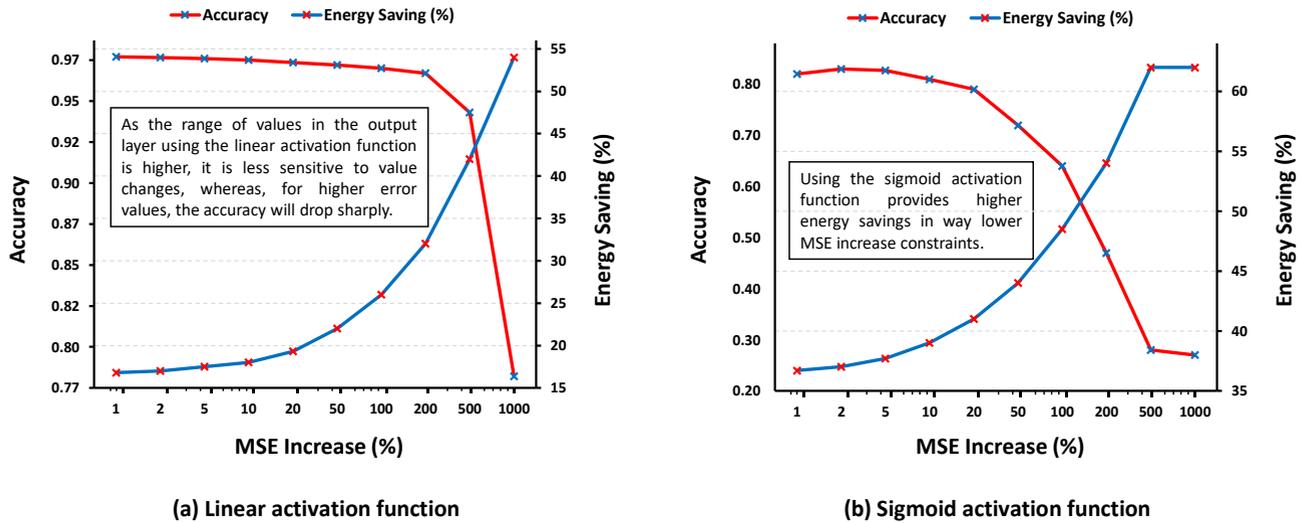

**FIGURE 13.** Accuracy drops and energy saving (secondary axis) of the studied FC network under the different MSE increment values, using a) linear and b) sigmoid activation functions.

In this figure, each row shows the determined operating voltage levels of all neurons distinguished by their number in the horizontal axis, e.g., a row shown by the red dotted box represents the operating voltage levels of all neurons of the FC DNN for the MSE increase upper bound of 100%. Now, based on obtained voltage values for a given quality constraint, to verify that the user constraints are met and the accuracy level is acceptable, a Gaussian noise with mean and variance determined from the statistical analysis has been injected into the TensorFlow model, and accuracy values have been calculated for all the constraints.

FIGURE 13 shows the accuracy drop of the NN model for the different MSE increase upper bounds, along with the amount of energy saving shown in the secondary axis. Based on the results, by increasing the MSE to 200% of the nominal value of the NN model that has been acquired using the test dataset, the proposed framework achieved 32% lower energy consumption only for 0.6% accuracy loss when the linear activation function was used. The same energy saving for the sigmoid activation function is achievable by choosing a lower MSE increment upper bound (e.g., 0.1%). Note that for the sigmoid activation function, the output of a neuron between 0 and 1, output MSEs are relatively small compared to Linear or ReLU activation functions.

Note that the activation function may affect the performance of the neural network. Mathematical expression, computational complexity [33], and the average processing time of each examined activation function calculated in our simulations are shown in Table 3, where the computational cost refers to the required computational resources for each function. As shown in the last column of this table, the ReLU is faster than TanH and Sigmoid due to its simple mathematical expression and efficient computation cost.

FIGURE 14 shows the accuracy and energy saving (secondary axis) of the LeNet-5 and ResNet-50 networks trained with the MNIST and CIFAR-10 datasets, respectively, under the different MSE increment values. For the studied range of MSE increment between 1% to 1000%, the LeNet-5 achieved, on average, 22% energy saving and an accuracy of 0.76. The minimum energy saving also belongs to the MSE=1%, where the accuracy of the network is 0.99 and energy saving is 6%. Also, before the accuracy dropping of the LeNet-5 below 90% (the green dotted line in FIGURE 14.a), the maximum energy saving is 18% and the accuracy is 0.92 (see pointer ⑤ in FIGURE 14.a). Moreover, the accuracy of this model drops below 0.8 for the MSE increment of more than 100% (see pointer ⑥ in FIGURE 14.a).

For the ResNet-50, under the same range of MSE increment, energy saving and accuracy are, on average, 35% and 0.66, respectively. For this network, the minimum energy saving is 13% for an accuracy of 0.92 (see pointer ⑦ in FIGURE 14.b). However, the accuracy of this model drops below 0.8 for the MSE increment of 10% (see pointer ⑧ in FIGURE 14.b).

TABLE 3 COMPUTATION COMPLEXITY AND AVERAGE PROCESSING TIME OF THE DIFFERENT ACTIVATION FUNCTIONS.

| Activation Function | Mathematical Expression | Computation Complexity [33] | Avg. Processing Time (S) |
|---|---|---|---|
| ReLU | $f(x) = \max(0, x)$ $= \begin{cases} x_i, & x > 0 \\ 0, & x \leq 0 \end{cases}$ | $O(1)$ | 1.12 |
| TanH | $f(x)$ $= \left(\dfrac{exp^x - exp^{-x}}{exp^x + exp^{-x}}\right)$ | $O(n^{2.085})$ | 1.50 |
| Sigmoid | $f(x)$ $= \left(\dfrac{1}{1 + exp^{-x}}\right)$ | $O(n^{2.085})$ | 1.48 |

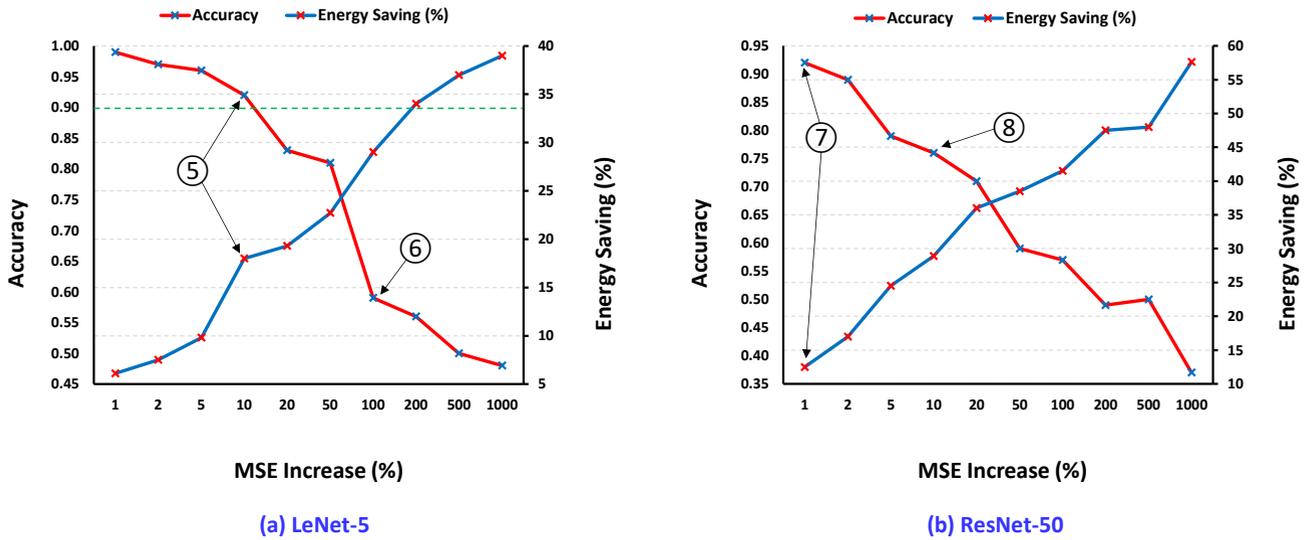

**(a) LeNet-5**

**(b) ResNet-50**

**FIGURE 14.** Network accuracy and energy saving (secondary axis) of the a) LeNet-5, and b) ResNet-50 networks trained with the MNIST and CIFAR-10 datasets, respectively, under the different MSE increment values.

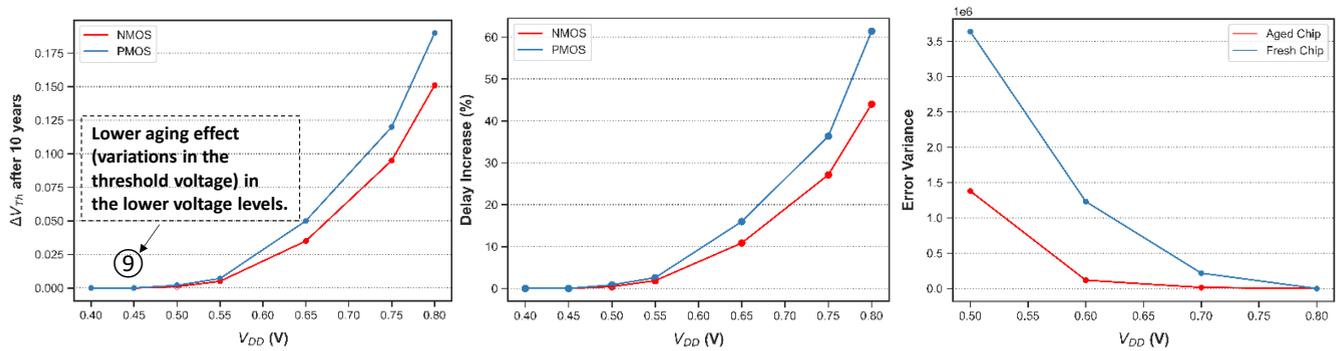

**FIGURE 15.** Effect of aging on the (a) threshold voltage, (b) path delay, and (c) error variance of a PE operating on various overscaled voltages, where the nominal voltage is 0.8V

### C. Aging Effect

To investigate the aging effect on X-TPU, (1) has been used to calculate threshold voltage changes after ten years of operation for the different studied overscaled voltages. The results of this investigation have been plotted in FIGURE 15 (a). Based on the results, the increment of threshold voltage for $V_{DD} = 0.8$ is 23.7% for PMOS (19% for NMOS), as this value is only 0.21% for $V_{DD} = 0.5$ for PMOS (0.2% for NMOS). Also, FIGURE 15 (b) shows the changes in path delay of transistors due to threshold voltage increase calculated using (3). Increasing path delay leads to increasing timing errors. However, to prevent these timing errors, the clock period is increased proportionally [8][15].

To analyze the effects of aging on the quality metrics of the X-TPU, we set the clock period of the circuit running at 0.8V after ten years of operation as the base clock time and simulated other operating voltages with this clock time. To this end, we used an in-house SDF modification tool to increase the amount of path delay of the SDF files of the synthesized PE according to FIGURE 15 (b). The results of the simulation have been plotted in FIGURE 15 (c). The clock period of a circuit is directly influenced by the critical path delay at the nominal voltage. However, by utilizing a lower $V_{DD}$, we can mitigate the aging effect and minimize the incremental delay of the gates over time (see pointer ⑨ in FIGURE 15) [8]. In our architecture, employing multiple voltages during circuit operation reduces overall aging because the circuit spends less time operating at the exact voltage. Additionally, higher voltages accelerate the rate of aging, resulting in increased circuit delay. This delay increment, along with the subsequent increase in the clock period, reduces the likelihood and severity of timing errors when operating at lower voltages. FIGURE 15 (c) represents the worst-case scenario in which we assume that there is a processing element (PE) in the X-TPU that consistently operates at the exact voltage. Alternatively, we can consider a uniform probability distribution of operating voltages. In this case, the delay increase in a PE would be the average delay increase across all selected voltages, as depicted in FIGURE 15 (b). By comparing this to the exact operation mode, we observe that the circuit's lifetime increases by 12%.

## VI. Conclusion

In this paper, we proposed a VOS-based runtime accuracy adjustable TPU architecture (X-TPU), and a framework to utilize this architecture to achieve significant energy efficiency while retaining the user-defined quality constraints. In specific, by knowing the supported voltage levels and the trained model parameters, the proposed framework can determine the operating voltage of each TPU column (corresponding to a neuron in a NN model) by maintaining the accuracy drop in an acceptable range. Running a DNN with linear activation functions, X-TPU can achieve 32% energy saving by sacrificing only 0.6% accuracy, and with a sigmoid activation function can save up to 40% energy with at most 0.5% accuracy loss. Also, we showed that using lower operating voltages on X-TPU can improve its lifetime.

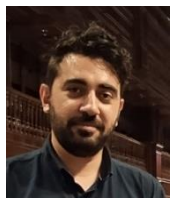

**Alireza Senobari** received his M.Sc. in computer architecture engineering from Tarbiat Modares University, Tehran, Iran, in 2022. He is currently a research assistant at Tarbiat Modares University in the Computer Architecture and Dependable systems Laboratory (CADS-Lab). His research interests include machine learning, neuromorphic computing, and fault-tolerant system design.

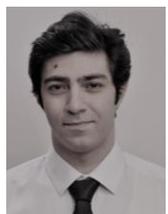

**Jafar Vafaei** received the B.Sc. degree from the University of Tabriz, Tabriz, Iran, in 2013, and the M.Sc. degree from University of Tehran, Tehran, Iran, in 2019, both in Electrical Engineering. He is currently a research assistant at Tarbiat Modares University, Tehran, Iran in the Computer Architecture and Dependable systems Laboratory (CADS-Lab). His research interests include low power digital designs and machine learning, reconfigurable computing, neuromorphic computing, and fault-tolerant system design.

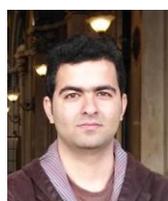

**Omid Akbari** received the B.Sc. degree from the University of Guilan, Rasht, Iran, in 2011, the M.Sc. degree from Iran University of Science and Technology, Tehran, Iran, in 2013, and the Ph.D. degree from the University of Tehran, Iran, in 2018, all in Electrical Engineering, Electronics - Digital Systems sub-discipline. He was a visiting researcher in the CARE-Tech Lab. at Vienna University of Technology (TU Wien), Austria, from Apr. to Oct. 2017, and a visiting research fellow under the Future Talent Guest Stay program at Technische Universität Darmstadt (TU Darmstadt), Germany, from Jul. to Sep. 2022.

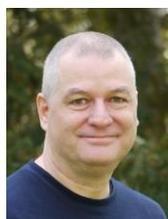

**Christian Hochberger** (Senior Member, IEEE) received the Diploma and Ph.D. degrees in computer science from TU Darmstadt, Darmstadt, Germany, in 1992 and 1998, respectively. After a short period as a freelance consultant, he became an Assistant Professor at the University of Rostock in 1999. He was an Associate Professor for embedded systems at the CS Department, TU Dresden, since 2003. He has been the Chair for Computer Systems at the EE Department, TU Darmstadt, since 2012. His research is focused on (re)configurable technology and design automation for such technology.

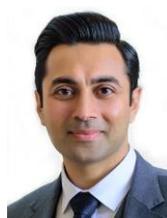

**Muhammad Shafique** (Senior Member, IEEE) received the Ph.D. degree in computer science from the Karlsruhe Institute of Technology (KIT), Germany, in 2011. In Oct.2016, he joined the Institute of Computer Engineering at the Faculty of Informatics, Technische Universität Wien (TU Wien), Vienna, Austria as a Full Professor of Computer Architecture and Robust, Energy-Efficient Technologies. Since Sep.2020, Dr. Shafique is with the New York University (NYU), where he is currently a Full Professor and the director of eBrain Lab at the NYU-Abu Dhabi in UAE, and a Global Network Professor at the Tandon School of Engineering, NYU-New York City in USA. He is also a Co-PI/Investigator in multiple NYUAD Centers. His research interests are in AI & machine learning hardware and system-level design, brain-inspired computing, quantum machine learning, cognitive autonomous systems, wearable healthcare, energy-efficient systems, robust computing, hardware security, emerging technologies, FPGAs, MPSoCs, and embedded systems.